\documentclass[conference]{IEEEtran}
\IEEEoverridecommandlockouts
\usepackage{cite}
\usepackage{amsmath,amssymb,amsfonts}
\usepackage{algorithmic}
\usepackage{graphicx}
\usepackage{textcomp}
\usepackage{xcolor}

\newcommand{\etal}{\textit{et al}. }
\newcommand{\ie}{\textit{i}.\textit{e}.}

\usepackage{geometry}
\def\BibTeX{{\rm B\kern-.05em{\sc i\kern-.025em b}\kern-.08em
    T\kern-.1667em\lower.7ex\hbox{E}\kern-.125emX}}
\begin{document}

\title{Ultra-sensitive Parity-Time Symmetry based Graphene FET (PTS-GFET) Sensors  \\
\thanks{*Shaolin Liao is the corresponding author.} }

\author{
 
\IEEEauthorblockN{Lu Ou}
\IEEEauthorblockA{\textit{College of Computer Science and Electronic Engineering} \\
\textit{Hunan University}\\
Changsha, Hunan, China \\
oulu9676@gmail.com} 

\and

\IEEEauthorblockN{Shaolin Liao*}
\IEEEauthorblockA{\textit{Department of Electrical and Computer Engineering} \\
\textit{Illinois Institute of Technology}\\
Chicago, IL, USA \\
sliao5@iit.edu; ORCID: 0000-0002-4432-3448}

}

\maketitle

\begin{abstract}
A novel ultra-sensitive Parity-Time symmetry based Graphene FET (PTS-GFET) sensor is studied for gas concentration detection. The PTS-GFET sensor effectively integrates the sensitivity of the PT symmetry around its Exceptional Point (EP) and the tunability of the GFET conductance. The change of GFET conductance with the gas concentration can be brought back to the EP of the PTS-GFET by tuning the gate voltage on the GFET. Thus, the applied gate voltage indicates the gas concentration. The minimum detectable gas concentration has been derived and estimated based on the experimental data, which shows that PTS-GFET can detect gas concentration below 50 ppb.
\end{abstract}


\section{Introduction}
Tunable Graphene Field Effect Transistor (GFET) can be used as resistance based sensitive sensors to measure ppb-level concentrations of gases such as NO$_2$ and NH$_3$ \cite{Yavari_2012}. However, working at low-frequency is prone to the $1/f$ noise, which reduces the sensitivity of the GFET sensors. 

Sensors working at the electromagnetic frequency \cite{liao_image_2006, liao_near-field_2006, shaolin_liao_new_2005, liao_fast_2006, liao_cylindrical_2006, liao_beam-shaping_2007, liao_fast_2007, liao_validity_2007, liao_high-efficiency_2008, liao_four-frequency_2009, vernon_high-power_2015, liao_multi-frequency_2008, liao_fast_2007-1, liao_sub-thz_2007, liao_miter_2009, liao_fast_2009, liao_efficient_2011, liao_spectral-domain_2019, liao_high_efficiency_2008, chan_single-pixel_2008, liao_sub_thz_2007, Bakhtiari_2011, Liao_2013, liao_novel_2014, liao_four_frequency_2009, nachappa_gopalsami_passive_2012, s._d._babacan_compressive_2011, gopalsami_compressive_2011, n._gopalsami_compressive_2011-1} can eliminate the $1/f$ noise to improve the sensors' performance. In addition, bandpass filters of high quality factor $Q$ can be used to further reduce the out-of-band noise. 

What's more, the emerging Parity-Time (PT) symmetry phenomenon at the electromagnetic frequency \cite{Hajizadegan_2019} can be explored to build very sensitive sensors around its Exceptional Point (EP), where the gain and loss balance each other and the ideal frequency splitting becomes zero. 

In this paper, we study an ultra-sensitive PT symmetry based GFET (PTS-GFET) gas sensor by integrating the sensitivity of the PT symmetry around its EP and the tunability of the GFET. 

 \begin{figure*}[th]
\centerline{\includegraphics[width=1\textwidth]{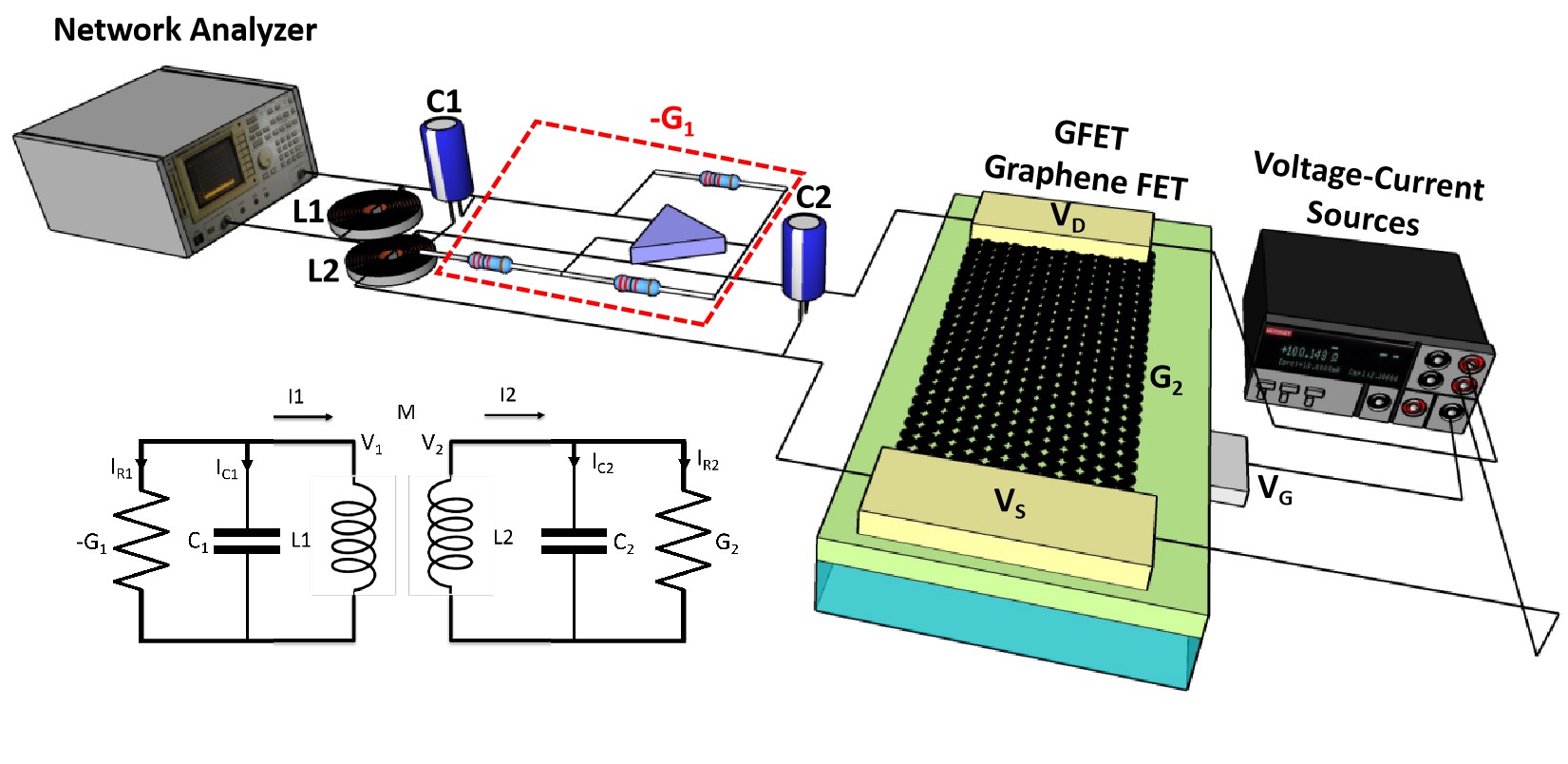}}
\caption{The PTS-GFET sensor concept and working principle: it consists of two RLC resonators, denoted by $(-G_1, L_1, C_1)$ and $(G_2, L_2, C_2)$. A VNA is used to obtain the reflection spectrum to find the EP of the PTS-GFET when the conductance of the GFET is tuned by the gate voltage through a voltage-current source, in an effort to cancel the effect of the gas concentration. Also, the bottom diagram shows the working principle of the PTS-GFET modeled as the coupled RLC resonators.}
\label{fig:problem}
\end{figure*}   

\section{The PTS-GFET Gas Sensor}
Fig. \ref{fig:problem} shows the concept of the PTS-GFET: it consists of a coupled RLC resonator pair, \ie, the $(-G_1, L_1, C_1)$ resonator on the reading side and the $(G_2, L_2, C_2)$ on the GFET sensing side. Idealy, the individual resonant frequencies of  $(-G_1, L_1, C_1)$ and $(G_2, L_2, C_2)$ are identical, \ie,  $\omega_1 = \omega_2 = 1/\sqrt{L_1 C_1}$. The negative conductance $-G_1$ is realized through a feedback amplifier \cite{Zhang_2019}. The conductance of the GFET $G_2$ changes with the gas concentration $c_{gas}$, as shown by the O-A path on the left plot of Fig. \ref{fig:IV}. To bring the PTS-GFET back to the EP, a gate voltage $V_G$ is applied to the GFET so that the changed GFET conductance $G_2$ is brought back to its origin, as shown by the A-B path on the left plot of Fig. \ref{fig:IV}. A Vector Network Analyzer (VNA) is used to scan the reflection spectrum to monitor that the PTS-GFET is actually brought back to the EP conditoin, \ie, $G_1 = - G_2$. Finally, the gate voltage $V_G$ will indicate the gas concentration $c_{gas}$.

\section{The Coupled RLC Coil Resonators}
\label{sec:formulation}
The PTS-GFET can be modeled as a coupled resonator pair. Each resonator can be modeled as a RLC resonant tank that consists of 3 components in parallel: a inductor with inductance L, a capacitor with capacitance C and a resistor R. 

\subsection{The Governing Equations System}
When a pair of  RLC resonant tanks are brought close to each other, they are coupled together through magnetic flux of the two inductive coils, which can be characterized by the mutual inductance $M$. The coupled resonant coils can be analyzed by the physical quantities of currents $i_1/i_{2}$ and voltages $v_{1}/v_{2}$ through the indicators of the two coupled coil resonators, 
\begin{align}\label{eqn:IV}
v_{1} = L_1 \frac{d i_1}{dt}  + M \frac{d i_{2}}{dt}, 
v_{2} = L_{2} \frac{d i_{2}}{dt}  + M \frac{d i_{1}}{dt}. 
\end{align}

The Kirchhoff Current Law (KCL) connects the currents through the inductor with inductance $L$, the capacitor with capacitance $C$ and the resistor with conductance $G$ as follows,
\begin{align}\label{eqn:KCL}
i_{1}  + C_1 \frac{dv_1}{dt} + G_1 v_1 = 0,   i_{2}  + C_2 \frac{dv_{2}}{dt} + G_{2}   v_{2}  = 0.   
\end{align} 

\subsection{The Quantum Hamiltonian}
In our recent paper \cite{Parity_Time_Liao_2020}, rigorous quantum Hamiltonian for the coupled RLC coil resonators system has been derived through twice basis transforms of the original basis. The first basis transform rotates the original basis such that off-diagonal terms of the governing matrix of the equation system of the coupled coil resonators reduces to constants. Then a second basis transform obtains the quantum Hamiltonian, including the diagonal effective complex frequencies and the off-diagonal coupling terms, together with the transformed basis. In particular, for identical and lossless coupled RLC resonators, \ie, $\omega_1 = \omega_2 = 1/\sqrt{L_1 C_1}$ and $G_1 = G_2 =0$,  the quantum Hamiltonian reads,
\begin{align}
 \mathcal{H}_0   = \begin{bmatrix}
\Omega&  1    \\
1 &  \Omega \\   
\end{bmatrix} ,
\end{align}
where $\Omega = \omega/\kappa$ is the normalized complex frequency and $\kappa$ is the effective coupling strength. 

For identical PT-symmetric coupled RLC resonators, \ie, $\omega_1 = \omega_2 = 1/\sqrt{L_1 C_1}$ and $G_2 = -G_1 =G$,  the effective quantum Hamiltonian can be expressed as
\begin{align}\label{eqn:Hamiltonian_PT}
 \mathcal{H}_{PT}   = \begin{bmatrix}
\Omega + j g &   1    \\
 1 &  \Omega - j g \\   
\end{bmatrix},
\end{align}
where $g=G/(2\kappa)$ is the normalized conductance.

If the coupled RLC resonators deviates in their own resonant frequencies and the gain doesn't balance the loss, \ie, $\omega_1 \neq \omega_2$ and $G_1 \neq -G_2$, the quantum Hamiltonian of Eq. (\ref{eqn:Hamiltonian_PT}) can be expressed as follows \cite{Mortensen_2018},
\begin{align}\label{eqn:Hamiltonian_PT_deviation}
 \mathcal{H}_{PT}'   = \begin{bmatrix}
\Omega + \Delta - j g' &   1    \\
1 &  \Omega -\Delta + j g \\   
\end{bmatrix},
\end{align}
where $\Delta = (\omega_1 - \omega_2)/(2 \kappa)$ is the normalized frequency deviation of the two RLC resonators. 

 \begin{figure*}[th]
\centerline{\includegraphics[width=1\textwidth]{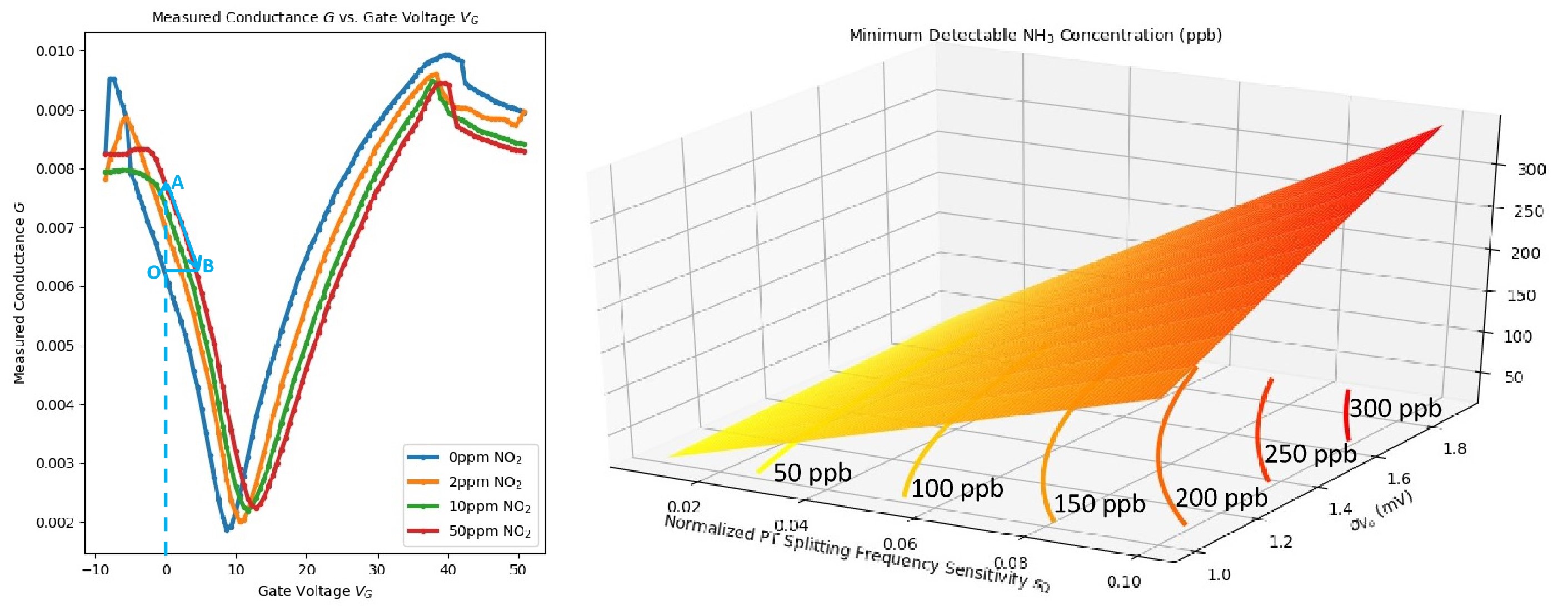}}
\caption{Conductance/resistance of GFET exposed to the NH$_3$ gas of different concentrations: left) Drain-source current $I$ vs. gate voltage $V_G$; and 2) the estimated minimum detectable gas concentration $c_{gas}$ calculated from the experimental data.}
\label{fig:IV}
\end{figure*}   

\subsection{The Eigenfrequencies}
The eigenfrequencies of the quantum Hamiltonian of Eq. (\ref{eqn:Hamiltonian_PT_deviation}) are given by,
\begin{gather}\label{eqn:eigenvalues}
\Omega_\pm = \Omega + j \frac{g-g'}{2} \pm \sqrt{1 - \left( \frac{g+g'}{2} +j \Delta \right)^2}.
\end{gather}

We are interested in the EP regime where $g' = 1;  g \sim 1$, which is realized through tuning the gate voltage $V_G$ of the GFET. For the normalized frequency deviation $\Delta \sim 0$, the eigenfrequencies of Eq. (\ref{eqn:eigenvalues}) can be approximated as follows,
 \begin{gather}\label{eqn:eigenvalues_2}
\Omega_\pm = \left(\Omega  \pm \sqrt{1 - \left( 1-  \Delta g/2 \right)^2}\right) - j \Delta g/2, \nonumber \\
 \sim \left(\Omega  \pm \sqrt{\Delta g} \right) - j \Delta g/2,
\end{gather}
where $\Delta g = 1-g$.

Assuming that the GFET conductance change $\delta g$ is much smaller than its Gaussian noise fluctuation standard deviation $\sigma$, \ie, $\delta g \ll \sigma$, it can be shown \cite{Mortensen_2018} that the normalized PT frequency splitting sensitivity is given by,
\begin{gather}\label{eqn:sensitivity_Omega}
s_\Omega \equiv   \sqrt{\frac{\pi \sigma}{2}}   = \sqrt{\frac{\pi \sigma}{2}}  \frac{\partial \langle \Omega_+ - \Omega_- \rangle}{\partial \Delta g}    \sim   \frac{\Delta g}{ \sigma }.
\end{gather}

\section{The Tunable GFET}
It is well-known that the conductance of the GFET $G$ can be tuned by applying the gate voltage $V_G$. In practical experimental environment, due to fluctuation of the gate voltage $\delta V_G$, the GFET's normalized noisy conductance $\delta g = \delta G/(2\kappa)$ given by,
\begin{gather}
\delta g = \frac{\delta G}{2\kappa} =  \frac{\partial g}{\partial V_G}  \delta V_G,
\end{gather}
from which the standard deviation of the normalized noisy conductance $\delta g$, denoted as $\sigma$, is obtained as follows,
\begin{gather}\label{eqn:standard_deviation_dg}
\sigma =   \frac{\partial g}{\partial V_G} \sigma_{V_G},
\end{gather}
where $\sigma_{V_G}$ is the standard deviation of the gate voltage $V_G$.

\section{The Minimum Detectable Gas Concentration}
The change of the graphene conductance $\Delta g$ due to the change of the gas concentration $\Delta c$ is given by,
\begin{gather}\label{eqn:gas_concentration}
\Delta g = \frac{\partial g}{\partial c_{gas}}  \Delta c_{gas}.
\end{gather}

Substituting Eq. (\ref{eqn:standard_deviation_dg}) and Eq. (\ref{eqn:gas_concentration}) into Eq. (\ref{eqn:sensitivity_Omega}), the normalized PT frequency splitting sensitivity due to the gate voltage fluctuation $\delta V_G$ is obtained,
\begin{gather}\label{eqn:sensitivity_Omega_dVG}
s_\Omega  \sim   \frac{\frac{\partial g}{\partial c_{gas}} }{\frac{\partial g}{\partial V_G} \sigma_{V_G} } \Delta c_{gas},
\end{gather}
from which the minimum detectable gas concentration is given by,
\begin{gather}\label{eqn:minimum_gas}
\Delta c_{gas} \sim   \frac{ \frac{\partial g}{\partial V_G} \sigma_{V_G} }{\frac{\partial g}{\partial c_{gas}} }  s_\Omega.
\end{gather}

We have estimated the minimum detectable gas concentration $c_{gas}$ based on the experimental data: 1) the change of GFET conductance with respect to the gate voltage $\partial g/\partial V_G$ is obtained from the measured GFET conductance vs. the gate voltage $V_G$, as shown on the left plot of Fig. \ref{fig:IV}; and 2) the change of the GFET conductance vs. the gas concentration $c_{gas}$ of NO$_2$ is obtained from the literature \cite{Yavari_2012}. Finally, the estimated minimum detectable gas concentration $c_{gas}$ for different gate voltage noise standard deviation $\sigma_{V_G}$ and normalized PT splitting frequency deviation $s_\Omega$ is shown on the right plot of Fig. \ref{fig:IV}. It can be seen that the minimum detectable gas concentration $c_{gas} < 50$ ppb when the gate voltage noisy deviation $\sigma_{V_G} = 1$mV and the normalized PT splitting frequency sensitivity $s_\Omega < 0.02$.

\section{Conclusion}
In this paper, an ultra-sensitive PTS-GFET sensor has been studied for gas concentration detection. The PTS-GFET effectively integrates the sensitivity of the PT symmetry phenomenon and the tunability of the GFET to form an ultra-sensitive sensor. Minimum detectable gas concentration has been analyzed. A minimum detectable gas concentration of $<50$ ppb has been estimated from the experimental data, showing that the PTS-GFET sensor is very promising in detecting tiny change of GFET conductance induced by physical parameters such as gas concentration, solute concentration in liquid and ultraviolet photons. 


\end{document}